\documentclass[useAMS,usenatbib]{mn2e}
\usepackage{graphicx}
\title[On the trigger mechanisms for SGR giant flares]{On the trigger mechanisms for SGR giant flares}

\author[R. Gill and J.S. Heyl]{Ramandeep Gill\thanks{E-mail: rsgill@phas.ubc.ca} and Jeremy S. Heyl\thanks{E-mail: heyl@phas.ubc.ca; Canada Research Chair} \\
Department of Physics and Astronomy, University of British Columbia, 6224 Agricultural Road,\\ Vancouver, British Columbia, Canada V6T 1Z1}
\begin{document}

\date{Accepted ---. Received ---; in original form ---}

\pagerange{\pageref{firstpage}--\pageref{lastpage}} \pubyear{2010}

\maketitle

\label{firstpage}

\begin{abstract}
We examine two trigger mechanisms, one internal and the other external to the neutron star, 
that give rise to the intense soft gamma-ray repeater (SGR) giant flares. So far, three giant flares 
have been observed from the three out of the seven confirmed SGRs on March 5, 1979, August 27, 1998, 
and December 27, 2004. The last two events were found to be much more powerful than the first, and 
both showcased the existence of a precursor, that we show to have had initiated the main flare. In 
the internal mechanism, we propose that the strongly wound up poloidal magnetic field develops 
tangential discontinuities and dissipates its torsional energy in heating the crust. The timescale 
for the instability to develop coincides with the duration of the quiescent state that followed 
the precursor. Alternatively, we develop a reconnection model based on the hypothesis 
that shearing motion of the footpoints causes the materialization of a Sweet-Parker current layer 
in the magnetosphere. The thinning of this macroscopic layer due to the development of an embedded 
super-hot turbulent current layer switches on the impulsive Hall reconnection, which powers the 
giant flare. Again, we show that the thinning time is on the order of the preflare quiescent time. 
This model naturally explains the origin of the observed nonthermal radiation during the flares, as 
well as the post flare radio afterglows.
\end{abstract}


\section{Introduction}
The soft gamma-ray repeaters (SGRs) showcase flux variability on many different timescales. 
The quiescent state, with persistent X-ray emission ($L_X\sim10^{35}\mbox{ erg s}^{-1}$), 
punctuated by numerous sporadic short bursts of gamma-rays, with peak luminosities 
up to $\sim 10^{42}\mbox{ erg s}^{-1}$ and typical duration in the range $\sim 0.01 - 1$ s, 
mark the defining characteristics of SGRs (see \citealt{Mereghetti2008} and 
\citealt{WoodsThompson2006} for a review). Out of the seven confirmed 
SGR sources, SGR 0525-66, SGR 1806-20, SGR 1900+14, SGR 1627-41, SGR 1150-5418, 
SGR 0418+5729, SGR 0501+4516 (with the last three added recently to the SGR family; see 
\citealt{Kanekoetal2010,Horstetal2010,Kumaretal2010}), the first three have been known to emit giant flares. 
A rare phenomenon compared to the commonly occurring short bursts, the giant flares unleash a 
stupendous amount of energy ($\sim 10^{44}$ erg) in gamma-rays over a timescale of $\sim 0.2-0.5$ s in a 
fast rising initial peak. The initial high energy burst is followed by a long ($\sim 200-400$ s), 
exponentially decaying pulsating tail of hard X-ray emission, the period of which 
coincides with that of the rotation of the neutron star (NS). Additionally, 
intermediate strength but rare outbursts lasting for 
few tens of seconds have been observed in the case of SGR 1900+14.

The first extremely energetic giant flare from a recurrent gamma-ray source, SGR 0525-66, was 
detected on March 5, 1979 by the gamma-ray burst detector aboard the Venera 11 \& 12 space 
probes and the nine interplanetary spacecraft of the burst sensor network 
\citep{Mazetsetal1979,HelfandLong1979}. The position of the source was found to be coincident 
with the supernova remnant N49 at a distance of $\sim55$ kpc in the Large Magellanic Cloud. 
The flare consisted of a sharp rise ($\sim 15$ ms) 
to the peak gamma-ray luminosity, $L_{\gamma}\sim10^{44}\mbox{ erg s}^{-1}$, subsequently 
followed by an exponentially decaying tail with $L_{\gamma}\sim10^{42}\mbox{ erg s}^{-1}$. 
Remarkably, the initial burst only lasted for $\sim0.1$ s compared to the longer lasting 
($\sim100$ s) tail that pulsated with a period of $\sim8$ s \citep{Terrelletal1980}. The 
total emitted energy during the initial peak and the decaying tail amounted to an 
astonishing $\sim10^{44}$ erg.

An even more energetic flare was detected from SGR 1900+14 on August
27, 1998 by a multitude of space telescopes in the direction of a
Galactic supernova remnant G42.8+0.6 \citep{Hurleyetal1999a}, making it
the second exceptionally energetic event detected in the past century
from a recurrent gamma-ray source. The burst had properties similar to
that of the March 5 event, with a short ($<4$ ms) rise time to the
main peak that lasted for $\sim 1$ s and then decayed into a pulsating
tail with a period identical to the rotation period of the NS of 5.16
s. The flare had a much harder energy spectrum compared to the March 5
event \citep{Ferocietal1999}, with peak luminosity in excess of
$\sim4\times10^{44}\mbox{ erg s}^{-1}$, assuming a distance of
$\sim10$ kpc to the source. The total energy unleashed in this 
outburst amounted to $\sim10^{44}$~erg in hard X-rays and gamma-rays \citep{Mazetsetal1999}.

Finally, on December 27, 2004 the most energetic outburst ever to be detected came from 
SGR 1806-20 \citep{Hurleyetal2005}, a Galactic source that was found to have a 
possible association with a compact stellar cluster at a distance of $\sim15$ 
kpc \citep{CorbelEikenberry2004}. The initial spike had a much shorter rise time ($\leq1$ ms) 
to the peak luminosity of $\sim2\times10^{47}\mbox{ erg s}^{-1}$ which persisted 
for a mere $\sim0.2$ s. Like other giant flares, a hard X-ray tail, of duration $\sim380$ s, 
followed the main spike 
pulsating at a period of 7.56 s. The total energies emitted during the initial spike and the 
harmonic tail are $\sim4\times10^{46}$ erg and $\sim10^{44}$ erg, respectively. 
\subsection{The Precursor}
\citet{Hurleyetal2005} reported the detection of a $\sim1$ s long precursor that was observed 
142 s before the main flare of December 27. A similar event, lasting a mere 
0.05 s \citep{Ibrahimetal2001}, was observed only 0.4 s prior to 
the August 27 giant flare \citep{Hurleyetal1999a,Ferocietal2001}, albeit at softer 
energies ($15-50$ keV, \citealt{Mazetsetal1999}); 
a non detection at harder energies ($40-700$ keV) was 
reported in \citet{Ferocietal1999}. Such a precursor was not detected at all for the March 
5 flare for which the detectors at the time had no sensitivity below $\sim50$ keV, which 
suggests that a softer precursor, if there indeed was one, may have gone unnoticed. 
Unlike the August 27 precursor, which was short and weak and for which no 
spectrum could be obtained \citep{Ibrahimetal2001}, the relatively longer lasting December 27 
precursor had a thermal blackbody spectrum with $kT\approx10.4$ keV \citep{Boggsetal2007}. 
In comparison to the common short SGR bursts, that typically last for $\sim$ 0.1 s and have 
sharply peaked pulse morphologies, the December 27 precursor was not only longer in duration 
but also had a nearly flat light curve. Nevertheless, the burst packed an energy $\sim 3.8\times10^{41}$ 
erg which is comparable to that of the short SGR bursts. 
The possible causal connection of the precursors to the giant flares in both cases indicates that they may have 
acted as a final trigger \citep{Hurleyetal1999a,Boggsetal2007}. A strong case for the 
causal connection of the precursor to the giant flare in both events can be established on statistical grounds. 
For example, SGR 1900+14 emitted a total of 50 bursts during its reactivation between May 26 and August 27, 1998 
following a long dormant phase lasting almost 6 years \citep{Hurleyetal1999b}. Here we are only interested in the 
burst history immediately prior to the Aug 27 event as this time period is indicative of the hightened activity 
that concluded with the giant flare. From these burst statistics, the rate of short bursts of typical duration 
$\sim$0.1 s is $\sim6\times10^{-6}\mbox{ s}^{-1}$, which then yields a null hypothesis probability of 
$\sim2.4\times10^{-6}$ for the August 27 precursor. Additionally, we find a null hypothesis probability 
of $\sim8.6\times10^{-4}$ in the case of the December 27 precursor, assuming similar burst rates. 
Although the magnetar model (particularly the phenomenological models developed in 
\citealt{ThompsonDuncan1995}, hereafter TD95, and \citealt{ThompsonDuncan2001}, hereafter TD01), 
as we discuss below, offers plausible explanations for the occurrence of 
short bursts and giant flares, the connection between the precursor and the main flare has remained 
unknown. In the event the precursor indeed acted as a trigger to the main flare, it is of 
fundamental significance that the association between the two events is understood.

As magnetars, SGRs are endowed with extremely large magnetic fields
with $B\sim 10^2 B_{\rm QED}$, 
where $B_{\rm QED}=4.4\times10^{13}$ G is the quantum critical field, and all the energetic phenomena 
discussed above are ascribed to such high fields (TD95). 
In the the TD95 model, the short bursts result due to sudden cracking of the crust as it fails 
to withstand the building stresses caused by the motion of the magnetic footpoints. The slippage 
of the crust, as a result, injects Alfv\'en waves into the external magnetic field lines, that 
subsequently damp to higher wavenumbers, and ultimately dissipate into a trapped thermal pair 
plasma. Such a mechanism may not be invoked for the giant flares due to energy requirements. 
Alternatively, a large-scale interchange instability \citep{Moffatt1985}, driven by the diffusion 
of the internal magnetic field, in combination with a magnetic reconnection event can power the 
giant flares. The plausibility of these mechanisms is well supported by the observed energetics 
of the bursts and the associated timescales. Nevertheless, a clear description of the reconnection 
process, which indubitably serves as one of the most efficient mechanisms to convert magnetic 
energy into heat and particle acceleration, has not been forthcoming. Furthermore, an alternative 
mechanism, motivated by the coronal heating problem in the solar case, can be formulated to give 
a reasonable explanation for the association of the precursor and the main flare.

In this paper, we propose two possible trigger mechanisms for the SGR giant flares - one internal 
and the other  external to the NS. As we argue, either of the two trigger mechanisms can 
initiate the main hyperflare. In the following discussion, we calculate model parameters 
for the December 27 event, however, the analysis is similar for the other two events. We start 
with a discussion of the internal trigger in the next 
section, followed by that of the external trigger in Section 3. The discussion regarding 
some of the observed characteristics of the flares that our model can account for is presented 
in Section 4.
\section{Internal trigger}
In the magnetar model, the magnetic field in the interior of SGRs is considered to be strongly wound up 
which then generates a strong toroidal field component, possibly even larger than the poloidal component 
(TD01). The relative strengths of the poloidal and toriodal magnetic field components have 
been quantified by constructing relativistic models of NSs and testing the stability of axisymmetric fields 
by \citet{LanderJones2009} and \citet{Ciolfietal2009}. Both studies arrive at the conclusion that the amplitude of the 
two field components may be comparable but the total magnetic energy is dominated by the poloidal component 
as the toroidal component is non-vanishing only in the interior, with $E_{\rm B,tor}/E_B\leq10$. However, 
in another study \citet{Braithwaite2009} arrived at a somewhat different conclusion where he found a significant 
enhancement in the toriodal component to sustain a stable magnetic field configuration, with 
$0.20\leq E_{\rm B,tor}/E_B\leq0.95$. In the interior, the twisted flux bundle, composed of several flux tubes, 
can be envisioned to stretch from one magnetic pole to the other along the symmetry axis of the 
dipole field that is external to the star. It has been shown by \citet{Parker1983b,Parker1983a} that 
any tightly wound flux bundle is unstable to dynamical nonequilibrium, and will dissipate 
its torsional energy as heat due to internal neutral point reconnection. Although Parker had provided such a 
solution to the long standing problem of coronal heating in the solar case, with a few exceptions, the same applies to the 
case of magnetars as the arguments are very general. In the case of 
the Sun, flux tubes are stochastically shuffled and wrapped around each other due to convective motions 
in the photosphere. Unlike in the Sun, where the flux tube footpoints are free to move in the photospheric 
layer, the footpoints are pinned to the rigid crust in NSs. Nevertheless, for exceptionally high magnetic 
fields ($B>10^{15}$ G) the crust responds plastically (TD01), and any moderate footpoint motion can still occur. 
It is understood that this is only true to the point where the crustal stresses are below some threshold, which 
depends on the composition. Thus, as the imposed strain exceeds some critical value, the crust will 
yield abruptly \citep{HorowitzKadau2009}, but may not fracture \citep{Jones2003}.  

Parker's solution is at best qualitative, however, it serves as a reasonably good starting point in the context 
of the present case. As we have noted earlier, the precursor may be causally connected to the main flare, and so 
can be argued to act as a trigger in the following manner. Immediately after the precursor the internal field 
evolves towards a new state of equilibrium. Since the crust has yielded to the built up stresses, 
and may deform plastically under magnetic pressure, some of the footpoints can now move liberally. 
Understandably, the turbulent dynamics, due to the high Reynolds number \citep{Peraltaetal2006}, 
of the internal fluid in 
response to the burst translates into chaotic motion of the footpoints. As a result, the flux tubes are 
wrapped around each other in a random fashion. Current sheets then inevitably form leading to reconnection 
followed by violent relaxation of the twisted flux bundle. The heat flux resulting from the dissipation of 
the torsional energy of the flux bundle is given by \citet{Parker1983b},
\begin{equation}
P = \left(\frac{B^2v^2\tau}{4\pi L}\right)
\label{eq:ParkerPower}
\end{equation}
where $B$ is the strength of the internal magnetic field, $v$ is the footpoint displacement velocity, 
and $L$ is the length scale of the flux tubes. Here $\tau$ is the timescale over which accumulation 
of energy by the random shuffling and wrapping of flux tubes occurs until some critical moment, 
after which neutral point reconnection becomes explosive. Having knowledge of the burst energetics, 
equation (\ref{eq:ParkerPower}) can be solved for $\tau$
\begin{equation}
\tau \sim 142\ B_{15}^{-2}L_6^{-1}E_{46}T_{0.125}^{-1}\left(\frac{v}{8.4\times10^3\mbox{ cm s}^{-1}}\right)^{-2}\mbox{ s}
\label{eq:ParkerTime}
\end{equation}
where we have used the event of December 27 as an example, with internal field strength measured in 
units of $10^{15}$ G, flux tube length scales in $10^6$ cm, the total energy of the flare in 
$10^{46}$ erg, and the timescale of the initial spike in 0.125 s (RHESSI PD time resolution). 
It is clear from equation (\ref{eq:ParkerTime}) 
that the preflare quiescent time scales linearly with the total energy emitted in the initial 
spike but is inversely proportional to the internal magnetic field strength: 
$\tau \propto E_{\rm spike}B_{\rm in}^{-2}$. Following TD95, we have assumed 
that almost all of the energy of the flare was emitted in the initial transient phase during which 
the lightcurve rose to its maximum. Additionally, we find that the footpoints are displaced 
at a rate of few tens of meters per second, which is a reasonable estimate considering the fact 
that it is insignificant in comparison to the core Alfv\'en velocity $V_A\sim10^7\mbox{ cm s}^{-1}$.

A noteworthy point is that in regards to the burst energetics there is 
nothing special about the precursor when compared to the common SGR bursts, other than that it 
occurs at the most opportune time when the internal field undergoes a substantial reconfiguration. 
The mechanism outlined above is activated after every SGR burst after which significant footpoint 
motion ensues. However, whether the entanglement of flux tubes is sufficient to reach a critical state 
such that an explosive release of energy can occur depends on the evolution of the internal field 
configuration.

Alternatively, the twisted flux bundle can become unstable to a resistive instability, such as 
the tearing mode. The resistivity here is provided by the turbulent motion of the highly 
conductive fluid which is in a state of nonequilibrium immediately after the precursor. The 
growth time of the tearing mode instability is given by the geometric mean of the Alfv\'en time, say 
in the core, and the resistive timescale
\begin{eqnarray}
\tau & = & (t_At_R)^{1/2} \\
& = & \left(\frac{4\pi\sigma L^3}{V_Ac^2}\right)^{1/2} \\
& \sim & 142\ L_6^{3/2}\left(\frac{V_A}{10^7\mbox{ cm s}^{-1}}\right)^{-1/2}\left(\frac{\sigma}{10^{13}\mbox{ s}^{-1}}\right)^{1/2}\mbox{ s} 
\end{eqnarray}
where $\sigma$ is not the electrical conductivity, but corresponds to the diffusivity of the 
turbulent fluid. In this case, the scaling for the preflare quiescent time becomes
\begin{equation}
\tau \propto E_{\rm spike}^{1/2}B_{\rm in}^{-3/2}
\end{equation}
where we have assumed that the twisted flux bundle occupied the entire 
internal region of the NS. 
\section{External trigger}
\begin{figure*}
\includegraphics{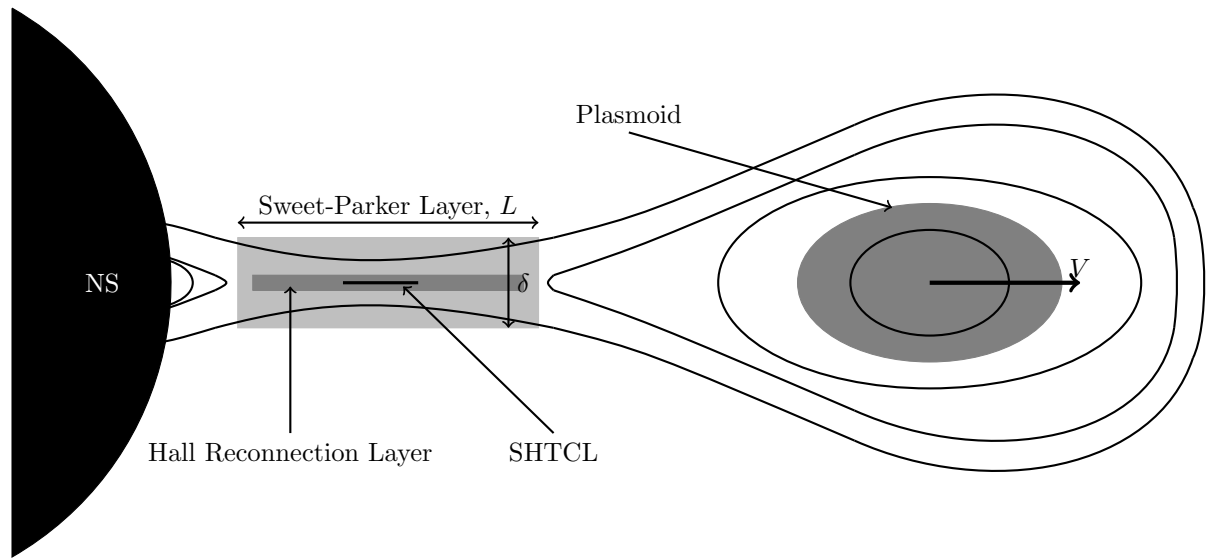}
\caption{This figure displays the setup of the different reconnecting current layers. 
The macroscopic Sweet-Parker layer with length $L\sim10^5$ cm and width $\delta\sim0.01$ 
cm is the largest of the three. This layer is then thinned down vertically as strong 
magnetic flux is convected into the dissipation region. The Hall reconnection layer, 
represented by the dark gray region, develops when $\delta$ becomes comparable to the 
ion-inertial length $d_i$. The system makes a transition from the slow to the impulsive 
reconnection and powers the main flare. The tiny region embedded inside the Sweet-Parker 
layer is the super-hot turbulent current layer, which aids in creating sufficient anomalous 
resistivity to facilitate the formation of the Sweet-Parker layer. The strongly accelerated 
plasma downstream of the reconnection layer is trapped inside magnetic flux lines and 
forms a plasmoid moving at some speed $V$. This plasmoid is then finally ejected during the 
initial spike when the external field undergoes a sudden relaxation (After \citealt{Lyutikov2006}).}
\label{fig:NSfig}
\end{figure*}
The notion that the giant flares are a purely magnetospheric phenomena appears very promising 
and requires further development. A magnetospheric reconnection model has become the favourite of many 
for two main reasons. First, it can easily explain the millisecond rise times of the explosive 
giant flares in terms of the Alfv\'en time of the inner magnetosphere, which for exceptionally low values 
of the plasma beta parameter is very small; $\tau_A\sim R_{\star}/c\sim 30 \mu$s. Second, 
the SGR giant flares have much in common with the extensively studied solar flares, for 
which reconnection models explaining nonthermal particle creation, plasma bulk motions, 
and gas heating have been developed over the last few decades \citep{Lyutikov2002}. 
The most powerful solar flares release an equally impressive amounts of 
energy $\sim 10^{32}$ erg, which is mainly divided into heating the plasma and radiation 
in multiple wavebands, for example $\gamma$-rays, X-rays, and radio. The impulsive rise in 
the soft X-ray emission to peak luminosity occurs over a timespan of few hundreds of seconds, 
which is then followed by a gradual decay lasting several hours \citep{PriestForbes2002}. 
In the magnetar model, because of the shearing of the magnetic footpoints 
caused by the unwinding of the internal field, a twist can be injected into the external 
magnetic field (TD95,TD01). Depending on how the crust responds to the stresses, either 
plastically or  rigidly, the gradual or sudden (in the event of a crustal fracture) 
transport of current from the interior creates a non-potential region in the magnetosphere 
where a reconnecting current layer can develop \citep{TLK2002,MikicLinker1994}. 
\citet{Lyutikov2003,Lyutikov2006} has explained the impulsive nature of the giant flares 
in terms of the tearing mode instability, which has a magnetospheric growth time of 
$\tau_{\rm tear}\sim 10$ ms.

Impulsiveness of the underlying magnetic reconnection mechanism explaining the origin 
of giant flares is a primary requirement. The tearing mode instability is quite befitting 
in that regard; however, it has not been shown to bear any dependence on the precursor, 
which as we argue, triggered the main hyperflare. Hall reconnection, which is another 
impulsive reconnection mechanism, has been completely ignored on the basis that it 
is unable to operate in a mass symmetric electron-positron pair plasma. 
Nevertheless, a mild baryon contamination may be enough to render it operational. 
Therefore, what is needed here is the synergy of two distinct
mechanisms --- a slow reconnection process, like the Sweet-Parker solution, that only dissipates 
magnetic energy at a much longer timescale \citep{Sweet1958,Parker1957}, 
and a fast process that is explosive, like Hall reconnection \citep{Bhattacharjeeetal1999}. 
To put this in the context of the December 27 event, we envision that immediately 
after the precursor a macroscopic current layer developed as a result of the 
sheared field lines. Then began the slow dissipation of magnetic field energy by 
Sweet-Parker reconnection, which continued throughout the quiescent state that 
followed the precursor. Finally, the transition to Hall reconnection resulted in 
the explosive release of energy (see figure~\ref{fig:NSfig}). We describe this process in more detail in the 
following section.
\subsection{Transition from Resistive to Collisionless Reconnection by Current Sheet Thinning}
The steady state reconnection process of Sweet and Parker is severely limited by its 
sensitivity to the size of the macroscopic dissipation region, such that the 
plasma inflow velocity is regulated by the aspect ratio of the current layer
\begin{equation}
v_i = \frac{\delta}{L}v_A
\label{eq:vin}
\end{equation}
where $\delta$ is the width and $L$ is the length of the dissipation region, 
with $\delta\ll L$ generally. The downstream plasma flow speed coincides with 
the Alfv\'en velocity, which in the magnetar case approaches the speed of light. 
The Sweet-Parker mechanism is a resistive reconnection process where the resistivity 
is either collisional or anomalous. It is understood that the electron-positron 
pair plasma pervading the inner magnetosphere is collisionless. Nevertheless, if 
enough ions are present in the dissipation region, as we show below, 
then a source of anomalous 
resistivity can be established. We argue that the energy released during the 
precursor was enough to heat the crust to a point where a baryon layer was 
evaporated into the magnetosphere. TD95 provide an upper limit to the mass 
of the baryon layer ablated during a burst by comparing the thermal energy 
of the burst to that of the potential energy of the mass layer
\begin{eqnarray}
\Delta M & \sim & \frac{E_{\rm th}R_{\star}}{GM_{\star}} \\
& \sim & 10^{17}\left(\frac{E_{\rm th}}{10^{38}\mbox{ erg}}\right)
\left(\frac{R_{\star}}{10^6\mbox{ cm}}\right)\left(\frac{M_{\star}}
{1.4\mbox{M}_{\odot}}\right)\mbox{ g}
\end{eqnarray}
where we have assumed a more conservative estimate of $E_{\rm th}$. Then, assuming 
that $\Delta M$ amount of baryonic mass, in the form of protons, was injected 
into the magnetospheric volume of $\sim R_{\star}^3$ yielding a baryon 
number density of
\begin{equation}
n_b \sim 6\times10^{22}\left(\frac{E_{\rm th}}{10^{38}\mbox{ erg}}\right)
\left(\frac{R_{\star}}{10^6\mbox{ cm}}\right)^{-2}\left(\frac{M_{\star}}
{1.4\mbox{M}_{\odot}}\right)\mbox{ cm}^{-3}
\label{eq:numberdensity}
\end{equation}
Even with the large amount of baryons, the magnetospheric plasma is still 
collisionless. The Spitzer resistivity for a quasi-neutral electron-ion 
plasma is only a function of the electron temperature $\propto T_e^{-3/2}$, 
which for electron 
temperatures as high as $\sim 10^8$ K yields a negligible resistivity. 
\subsubsection{Super-Hot Turbulent Current Layer}
For plasma temperatures higher than $T>3\times10^7$ K, the 
reconnecting current layer turns into a super-hot turbulent current 
layer (SHTCL), for which the theory has been well developed by and 
documented in (\citealt{Somov2006}, pp. 129-151). The anomalous resistivity in 
the current layer arises due to wave-particle interactions, where 
the ions interact with field fluctuations in the waves. As a result, 
the resistivity and other transport coefficients of the plasma are 
altered. The electrons are the current carriers and participate 
mainly in the heat conductive cooling of the SHTCL. The current 
layer is assumed to have been penetrated by a relatively weak 
transverse magnetic field component (transverse to the electric 
field in the current layer), where $B_\perp\ll B_0$ with 
$B_0$ as the strength of the external dipole field. In the two 
temperature model, where the electrons and ions are allowed to 
have dissimilar temperatures, the effective anomalous resistivity 
is generally a combination of two terms. One resulting from the 
ion-acoustic turbulence and the other from the ion-cyclotron 
turbulence. 
\begin{equation}
\eta_{\rm eff} = \eta_{\rm ia} + \eta_{\rm ic}
\end{equation}
In addition, each turbulent instability has two separate 
regimes -- marginal and saturated. The former applies when the wave-particle 
interactions are described by quasilinear equations, and the latter 
becomes important in the case of strong electric fields when the 
nonlinear contributions can no longer be ignored (see for e.g. 
\citealt{Somov1992} pp. 115-217 for a detailed description). For 
an equal temperature plasma ($T_e\sim T_i$), the saturated 
ion-cyclotron turbulent instability makes the dominant contribution 
to the effective resistivity. Thus, we ignore any other terms 
corresponding to the ion-acoustic instability. The effective resistivity 
in the present case is given as \citep{Somov2006}, depending on the 
dimensionless temperature parameter $\theta\equiv T_e/T_i$,
\begin{equation}
\eta_{\rm eff} = \frac{2m_e^{1/2}\pi^{1/4}}{ec^{1/2}m_p^{1/4}}
\left[\frac{(1+\theta^{-1})^{1/2}}{N^{1/4}(\theta)U_k(\theta)}
\right]\frac{(B_\perp E_0)^{1/2}}{B_0^{1/2}n_b^{3/4}}
\label{eq:effectiveEta}
\end{equation}
where
\begin{eqnarray}
N(\theta) & = & 1.75 + \frac{f(\theta)}{\sqrt{8}(1+\theta^{-1})^{3/2}} \\
f(\theta) & = & \frac{1}{4}\left(\frac{m_p}{m_e}\right)^{1/2}\mbox{  for  }
1\leq\theta\leq8.1 \\
U_k(\theta) & \sim & \mathcal{O}(1)\mbox{  for  }\theta\sim 1 \\
E_0 & = & \alpha B_0
\label{eq:frozenin}
\end{eqnarray}
$\alpha\equiv v_0/c$ is the effective reconnection rate determined 
by the inflow fluid velocity $v_0$ into the current layer, 
and the rest of the variables in equation (\ref{eq:effectiveEta}) retain their usual meaning. 
Equation (\ref{eq:frozenin}) conveys the frozen-in field condition. Next, we write the 
magnetic diffusivity of the plasma due to the effective anomalous resistivity
\begin{eqnarray}
\eta_{\rm diff} & = & \frac{\eta_{\rm eff} c^2}{4\pi} \\
& \simeq & 6\times10^{23}(\alpha B_\perp)^{1/2}n_0^{-3/4}
\end{eqnarray}

To calculate the inflow plasma velocity, we assume that the SHTCL is 
embedded in a macroscopic Sweet-Parker current layer. The primary role 
of the SHTCL is to provide enough resistivity in a collisionless plasma 
so that the magnetic field lines can diffuse through it and ultimately 
undergo reconnection. From equation (\ref{eq:vin}) we know that for a Sweet-Parker 
current layer the inflow fluid velocity is regulated by the aspect 
ratio of the current layer. The outflow velocity is limited by the 
speed of light. By expressing the width of the Sweet-Parker current 
layer in terms of the magnetic diffusivity, we find that the inflow 
velocity has to be on the order of $v_0\sim10^3\mbox{ cm s}^{-1}$ (so 
that $\alpha\ll1$), with the width of the layer given as
\begin{eqnarray}
\delta & \sim & \sqrt{\frac{\eta_{\rm diff} L}{c}} \\
& \sim & 0.01\mbox{ cm }\left(\frac{v_0}{10^3\mbox{cm s}^{-1}}\right)^{1/4}\left(\frac{B_\perp}
{10^{11}\mbox{ G}}\right)^{1/4} \\
& &\times\left(\frac{n_b}{6\times10^{22}\mbox{ cm}^{-3}}\right)^{-3/8}
\left(\frac{L}{10^5\mbox{ cm}}\right)^{1/2} \nonumber
\end{eqnarray}
where the transverse magnetic field is $B_\perp\sim10^{-3}B_0$, and 
$L$ is the length of the current layer. The size of the SHTCL 
can now be obtained from the following
\begin{eqnarray}
a & = & \frac{c}{e}\sqrt{\frac{m_e}{2\pi n_b}}\left[\sqrt{\frac{1+\theta^{-1}}
{N(\theta)}}\frac{1}{U_k(\theta)}\right] \\
& \sim & 2.5\times10^{-6}\left(\frac{n_b}{6\times10^{22}\mbox{ cm}^{-3}}
\right)^{-1/2}\mbox{ cm} \\
b & = & \frac{B_0}{h_0}\sqrt{\frac{2v_0}{B_\perp}}\left[\frac{\pi m_p
n_b}{N(\theta)}\right]^{1/4} \\
& \sim & 80\mbox{ cm }\left(\frac{R_{\star}}{10^6\mbox{ cm}}\right)\left(\frac{v_0}
{10^3\mbox{ cm s}^{-1}}\right)^{1/2} \\
& & \times\left(\frac{B_\perp}{10^{11}\mbox{ G}}\right)^{-1/2}\left(
\frac{n_b}{6\times10^{22}\mbox{ cm}^{-3}}\right)^{1/4} \nonumber
\end{eqnarray}
where $a$ and $b$ are, respectively, the half-width and the half-length of the 
SHTCL, and $h_0\sim B_0/R_{\star}$ is the magnetic field gradient in the 
vicinity of the current layer.
\subsubsection{Current Sheet Thinning}
The main flare is triggered when the transition is made from the steady state, 
slow reconnection process to an impulsive one. In the present case, Sweet-Parker 
reconnection makes a transition to Hall reconnection when the width of the 
current layer $\delta$ drops below the ion-inertial length $d_i$, where
\begin{eqnarray}
d_i & = & \frac{c}{\omega_{p,i}} = \frac{c}{e}\sqrt{\frac{m_p}{4\pi n_b}} \\
& \sim & 10^{-4} \left(\frac{n_b}{6\times10^{22}\mbox{ cm}^{-3}}
\right)^{-1/2}\mbox{ cm}
\end{eqnarray}
and $\omega_{p,i}$ is the non-relativistic ion plasma frequency. 
\citet{Cassaketal2005} show that for a given set of plasma parameters, 
the solution is bistable such that the slow Sweet-Parker solution 
can operate over long timescales, during which the system can accumulate 
energy, while the faster Hall solution starts to dominate as the resistivity 
is reduced below some critical value. Lowering the resistivity would naturally 
reduce the width of the current layer to the point where the system can access 
the Hall mechanism. Alternatively, as \citet{Cassaketal2006} argue, the same 
result can be achieved by thinning down the current layer by convecting in 
stronger magnetic fields into the dissipation region during Sweet-Parker 
reconnection. The critical field strength needed to thin the current layer 
is
\begin{eqnarray}
B_{c} & \sim & \sqrt{4\pi m_pn_b}\left(\frac{\eta_{\rm diff}}{d_i^2}L\right) \\
& \sim & 4\times10^{14}\mbox{ G }\left(\frac{n_b}{6\times10^{22}
\mbox{ cm}^{-3}}\right)^{3/4} \\
& & \times\left(\frac{v_0}{10^3\mbox{ cm s}^{-1}}
\right)^{1/2}\left(\frac{B_\perp}{10^{11}\mbox{ G}}\right)^{1/2}
\left(\frac{L}{10^5\mbox{ cm}}\right) \nonumber
\end{eqnarray}
Due to flux pile up outside the current layer, it can be argued 
that the system is able to achieve such high field strengths. The 
timescale for thinning down the current sheet until its width is 
comparable to the ion-inertial length is given as
\begin{eqnarray}
\tau_{\rm thin} & \sim & 2W_s\sqrt{\frac{L}{\eta_{\rm diff}c}\left(\frac{B_c}{B_0}
\right)} \\
& \sim & 130\mbox{ s }\left(\frac{W_s}{10^5\mbox{ cm}}\right)\left(
\frac{L}{10^5\mbox{ cm}}\right)\left(\frac{B_0}{10^{14}\mbox{ G}}
\right)^{-1/2} \\
& & \times\left(\frac{n_b}{6\times10^{22}\mbox{ cm}^{-3}}\right)^{3/4}
\nonumber
\end{eqnarray}
where $W_s$ is the magnetic shear length, that is the length scale over 
which the field lines are severely sheared. What we find here is that the 
thinning down time $\tau_{\rm thin}$ of the current layer from the Sweet-Parker width to the 
ion-inertial length, where Hall reconnection dominates, is on the 
order of the preflare quiescent time of $\sim 142$ s for the December 27 event. 
The scaling relation of the thinning down time in terms of the initial 
spike energy and the external magnetic field strength can be deduced to 
be the following
\begin{equation}
\tau_{\rm thin} \propto E_{\rm spike}^{2/3}B_0^{-11/6}
\label{eq:scalingthin}
\end{equation}
Again, we emphasize here that this same mechanism may operate after every 
SGR burst which is energetic enough to inject the requisite baryon number density, as 
calculated in equation (\ref{eq:numberdensity}), to facilitate the development of 
a Sweet-Parker current layer. However, this mechanism will fail if the twist injected 
into the magnetosphere by the unwinding of the internal field is not sufficient 
to create a tangential discontinuity at the first place. In that instance no current 
sheet will form. 
\subsubsection{Giant flare Submillisecond Rise Times}
In Hall reconnection, a multiscale dissipation region develops with characteristic spatial 
scales on the order of the ion and electron inertial lengths \citep{Shayetal2001}. 
Within a distance $d_i$ of the neutral X-line, the ions decouple from the electrons and 
are accelerated away at Alfv\'enic speeds in the direction perpendicular to that of the 
inflow. The electrons continue their motion towards the neutral line as they are frozen-in, 
and only decouple from the magnetic field when they are a distance $d_e$, the 
electron-inertial length, away from the neutral line. Within the ion-inertial 
region, the dynamics of the electrons are significantly influenced by the nonlinear 
whistler waves. Subsequently, the electrons too are accelerated away in an 
outflowing jet at Alfv\'enic speeds. The timescale associated to Hall reconnection 
then is in good accord with the rise times of giant flares \citep{Schwartzetal2005}, 
that is
\begin{equation}
\tau_{\rm Hall} \sim \frac{R_{\star}}{0.1 c} \sim 0.3\mbox{ ms}
\end{equation}
\section{Discussion}
In this paper, we present an internal and an external trigger mechanism for the 
SGR giant flares, where we strongly emphasize the causal connection of the precursor 
to the main flare. The quiescent state that follows the precursor has been argued, in 
our model, to be the time required for the particular instabilities to develop, along 
with the accumulation of energy just before the flare. The internal mechanism is based 
on the hypothesis that poloidal field component in the interior of the NS is strongly 
wound up. The solution is motivated by Parker's reasoning that such a twisted field 
would inevitably develop tangential discontinuities and dissipate its torsional energy 
as heat. The timescale for the accumulation of energy that is to be released in the 
main flare is on the order of the duration of the preflare quiescent state.

The external trigger mechanism makes use of the fact that a Sweet-Parker reconnection 
layer may develop between significantly sheared field lines if a source of resistivity 
is established. Such a source may be embedded inside the macroscopic Sweet-Parker 
layer in the form of a super-hot turbulent current layer. To make the reconnection 
process impulsive, we invoke the non-steady Hall reconnection which is switched on 
as the width of the Sweet-Parker layer is thinned down to the ion-inertial length. 
Again, the timescale over which the layer is thinned down roughly coincides with that 
of the preflare quiescent state duration.  

We have shown detailed calculations of the timescales for the December 27 event 
in particular. However, a similar analysis can also be carried out for the August 27 
event. For the internal mechanism, we find the timescale to be comparable to the 
observed preflare quiescent time with $\tau\sim0.4\ B_{15}^{-2}L_6^{-1}E_{44}T_1^{-1}
\left(\frac{v}{5.6\times10^4\mbox{ cm s}^{-1}}\right)^{-2}$ s, where we have assumed 
the same internal magnetic field strength and length of flux tubes. For the external 
mechanism, assuming $\Delta M\sim10^{15}$ g, since the precursor was short and weak, 
$W_s\sim2\times10^4$, and $L\sim5\times10^4$, we find $\tau_{\rm thin}\sim0.4$ s, 
$\delta\sim0.04$ cm, $a\sim2.5\times10^{-5}$ cm, $b\sim25$ cm, and 
$d_i\sim10^{-3}$ cm.

A significant nonthermal component, with an average power-law index of $\Gamma\sim2$ 
as in $E^{-\Gamma}$, 
was observed during the decaying phase of the flare in both the August 27 and 
December 27 events \citep{Ferocietal1999,Boggsetal2007}. In the magnetar model, the 
nonthermal emission originates much farther out from the star, almost at the light 
cylinder (TD01). At this distance, inverse Compton cooling by 
X-ray photons has been invoked to explain the nonthermal spectrum. Nonthermal particle 
generation is one of readily identified features of magnetic reconnection, especially 
in the case of Hall reconnection where outflow velocities approach the Alfv\'en 
speed of the medium. Such acceleration of high energy particles due to meandering-like 
orbits in the presence of strong electric fields has also been seen in 
particle-in-cell simulations \citep{ZenitaniHoshino2001}. Therefore, the Hall 
reconnection process that gives rise to the main flare can easily explain the 
origin of nonthermal particles.

\citet{Israeletal2005} and \citet{StrohmayerWatts2005} reported the detection of quasi-periodic 
oscillations (QPOs) in the burst spectra of the December 27 and August 27 events, 
respectively. QPOs in the December 27 event were detected at 92.5 Hz, and 18 and 30 Hz 
at, respectively, 170 s and $\sim200-300$ s after the initial spike. \citet{WattsStrohmayer2006} 
confirmed the detection of the first two QPOs and reported the 
presence of two additional QPOs at 26 and 626.5 Hz. Similarly, in the 
August 27 event, QPOs at 84 Hz, 53.5 Hz, 155.1 Hz, and 28 Hz (with lower significance) 
were detected at about a minute 
after the onset of the flare. Torsional oscillation of the NS crust appeared to be the 
natural explanation for the QPOs. However, \citet{Levin2006} argues that purely crustal 
oscillations rapidly lose their energy to an Alfv\'en continuum in the core by resonant 
absorption. He demonstrates that steady low frequency oscillations can be associated 
with MHD continuum turning points in the core \citep{Levin2007}, while others have reproduced the QPOs 
in toy models governed by global MHD-elastic modes of the NS \citep{Glampedakisetal2006,
Sotanietal2007}. Both trigger mechanisms presented in this paper can be linked to the 
initiation of such an oscillatory behavior, whether in the core or as a global mode of the 
star, by the realization that during the giant flare the global magnetic field of the 
star undergoes sudden magnetic relaxation (TD95). In the internal 
trigger, the sudden loss of helicity can be argued to be sufficient to launch Alfv\'en 
waves in the interior. 

On the other hand, although the external trigger, is not directly tied to the crust, 
however, a sudden relaxation of the internal toroidal field in the sense of 
the \citet{FlowersRuderman1977} instability can be realized. The loss of magnetic 
energy in the form of a plasmoid, which has been know to form during an eruptive 
flare \citep{MagaraShibata1997}, serves to relax both the sheared external dipole 
field and the twisted internal toroidal field. Since both fields thread through 
the entire star, a sudden relaxation during the initial spike can easily excite 
global elastic modes in the NS. The ejection of a plasmoid also naturally explains 
the origin of the radio afterglow observed for both the August 27 and the December 27 events 
(\citealt{Frailetal1999,Gaensleretal2005}; also see \citealt{Lyutikov2006} 
for afterglow geometry and parameters).

In our calculations, we have shown how the preflare quiescent time scales 
with the total energy released in the initial spike. In the internal trigger 
mechanism, we find that Parker's solution yields a linear dependence. Although, 
as we have remarked earlier, this mechanism is simple and elegant but based 
on qualitative arguments. Nevertheless, it does reproduce the observed 
result, that is the longer the preflare quiescent time the energetic the flare, 
if the internal magnetic field strengths for both NSs are 
assumed to be similar: $\tau_{\rm Aug}/\tau_{\rm Dec} \sim E_{\rm
  Aug}/E_{\rm Dec}$. 
On the other hand, for the scaling to reconcile with observations in the 
case of the tearing mode instability and the external trigger, either 
$E_{\rm Aug}/E_{\rm Dec} \ll 10^{-3}$ or $B_{1900+14}/B_{1806-20} \gg 10^{-1}$. Based 
on the measured $P$ and $\dot{P}$ values for SGR 1900+14 and SGR 1806-20, 
which suggest that $B_{1806-20}\sim3B_{1900+14}$, and with the revised 
distance estimate of $D\sim12-15$ kpc for SGR 1900+14 \citep{Vrbaetal2000}, 
neither condition may be satisfied. However, one should not ignore the fact 
that the external mechanism also depends on the size of the current layer 
and the field line shearing lengthscale. Therefore, the scaling relation 
may not be as simple as that argued in equation (\ref{eq:scalingthin}). In any case, 
with only two events it is premature to observe any trends regarding the 
preflare quiescent times and the burst energies. Future giant flares from 
SGRs will certainly improve our understanding of such correlations.
\section*{Acknowledgements}
We would like to thank the anonymous reviewer for his help in improving the quality 
of this paper. R.G. is supported by NSERC CGS-D3 scholarship. 
The Natural Sciences and Engineering
Research Council of Canada, Canadian Foundation for Innovation and the
British Columbia Knowledge Development Fund supported this work.
Correspondence and requests for materials should be addressed to
J.S.H. (heyl@phas.ubc.ca).  This research has made use of NASA's
Astrophysics Data System Bibliographic Services.

\label{lastpage}

\end{document}